# Single photon emission in WSe$_2$ up 160 K by quantum yield control

*Yue Luo[1,2], Na Liu[1,2], James C. Hone[3] and Stefan Strauf [1,2]\**

[1] Department of Physics, Stevens Institute of Technology, Hoboken, NJ 07030, USA

[2] Center for Quantum Science and Engineering, Stevens Institute of Technology, Hoboken, NJ 07030, USA

[3] Department of Mechanical Engineering, Columbia University, New York, NY 10027, USA

\*Address correspondence to: strauf@stevens.edu

**Abstract**

Two dimensional semiconductors hosting strain-induced quantum emitters offer unique abilities to achieve scalable architectures for deterministic coupling to nanocavities and waveguides that are required to enable chip-based quantum information processing technologies. A severe drawback remains that exciton emission from quantum emitters in WSe$_2$ quenches beyond 30 K, which requires cryogenic cooling. Here we demonstrate an approach to increase the temperature survival of exciton quantum emitters in WSe$_2$ that is based on maximizing the emitter quantum yield. Utilizing optimized material growth that leads to reduced density of nonradiative defects as well as coupling of the exciton emission to plasmonic nanocavities modes, we achieve average quantum yields up to 44%, thermal activation energies up to 92 meV, and single photon emission signatures up to temperatures of 160 K. At these values non-cryogenic cooling with thermo-electric chips becomes already feasible, while quantitative analysis shows that room temperature operation is within reach through active strain engineering.

**Introduction**

Solid-state quantum emitters that operate as true single photon sources or spin-based quantum bits are a critical resource to enable chip-based quantum information processing[1]. However, no current technology can create quantum emitters combining room-temperature operation with deterministic placement needed for coupling to cavities and other optical structures. For instance, color centers in diamond[2], silicon carbide[3,4], or hexagonal boron nitride[5–7] are well known to show quantum light signatures up to and above room temperature. The color center itself is a point defect that has a symmetry group and energy level structure distinct from the host crystal band structure, giving rise to energetically deeply localized electrons that are robust against temperature, with record high values for single photon emission from boron nitride quantum emitters up to 800 K [8]. However, spatial control of color centers with a precision of tens of nm as required for deterministic coupling to nanocavities remains difficult, since most color centers not only require a foreign atom but also activation through paring with a local lattice vacancy[9]. Zero-dimensional excitons, on the other hand, can be engineered using a confinement potential, typically in a heterostructure with a surrounding material of larger bandgap. In this way, single photon emission from CdSe [10] and GaN [11] quantum dots have been demonstrated at 200 K and even room temperature with sufficient exciton confinement through embedding in ZnSe or GaN nanowires[12,13]. Alternatively, individual molecules like single-walled carbon nanotubes offer the ability to trap excitons through sidewall-functionalization chemistry. While initial work on carbon nanotubes demonstrated single photon emission from disorder-localized excitons only at cryogenic temperatures[14–16], more recent work utilizes covalently bound oxygen or $sp^3$ defects that trap excitons into energetically deep states that are robust up to room temperature [17,18] However, spatial control of quantum dot or carbon nanotube-based emitters remains difficult.

Two-dimensional materials such as the transition metal dichalcogenide semiconductor tungsten diselenide ($WSe_2$) are a promising new host for zero-dimensional excitons. In planar monolayers these emitters are trapped into random locations[19–22], but recent work has shown that nanobubbles[23] or lithographically defined dielectric pillars[24,25] can confine emitters to local areas of high strain, thereby enabling spatially controlled quantum emitter formation. We have recently utilized this phenomenon to create arrays of up to 60 quantum emitters in $WSe_2$ coupled plasmonic nanocavities[26]. At demonstrated single photon emission rates up to 42 MHz under 78 MHz excitation the system approaches operation as a deterministic quantum light source, since already every second trigger pulse creates a single photon from the device. Despite these appealing properties of quantum emitters in $WSe_2$, their operation has been limited to cryogenic temperatures since the 0D exciton emission was shown to quickly quench beyond 30 K[19,27].

Here we demonstrate an approach to increase the temperature range of exciton quantum emitters that is based on maximizing the quantum yield (QY) of strain-induced excitons in monolayer $WSe_2$. Initially, these quantum emitters have a rather low quantum yield of typically only 1% due to thermally activated nonradiative (NR) recombination, limiting exciton emission to liquid helium temperatures. By combining advances in the material growth that reduces nonradiative (NR) defect centers with coupling to plasmonic nanocavities to enhance the radiative exciton recombination rate through the Purcell effect, we show that quantum light emission in monolayer $WSe_2$ can exceed 160 K for an average quantum yield of 44% and activation energy of 92 meV. In addition, we quantitatively studied temperature dependent exciton dephasing, bandgap energy and exciton intensity and predict the minimum requirement to enable room temperature operation by strain engineering.

**Results and discussion**

Thermal activation of optically pumped excitons out of quantum confined energy states can be described based on two-level rate equation analysis that includes radiative and NR recombination rates $\tau_{NR}$ and $\tau_R$, respectively[28]. Assuming $\tau_{NR}$ is thermally activated, either via level depopulation or by direct activation of NR centers in the material, one can derive the steady-state solution for the temperature-dependent intensity $I$ of the exciton recombination, which is given by $I = I_0 \left[1 + a\, exp(-E_A/kT)\right]^{-1}$, where $E_A$ is the thermal activation energy and $a$ describes the ratio $\tau_{NR}/\tau_R$ that is directly related to the QY of the host material[29]. Apparently, there are two ways to increase the survival of optical emission of strain-induced quantum emitters against increasing material temperature. On one hand one can attempt to create deeper localization potentials by applying larger local strain on the 2D material, which increases $E_A$. On the other hand, one can decrease the ratio of $\tau_{NR}/\tau_R$ by directly enhancing the material QY, which can be described as QY = $\gamma_R/(\gamma_R + \gamma_{NR})$, where $\gamma_R = 1/\tau_R$ and $\gamma_{NR} = 1/\tau_{NR}$. In our experiments we focus on improving the QY of the strain-induced quantum emitters to not only reach higher temperature operation but also overall brighter emission.

With the goal to realize higher QY and thus better temperature survival for strain-induced quantum emitters we first focus on an approach that drastically enhances $\tau_R$ by the Purcell effect. Following our previous work, quantum emitters in monolayer WSe$_2$ can be created in a spatially deterministic way using the sharp corners of Au nanocubes that induces strain at the cube corners and thus a localization potential resulting in quantum-dot like exciton emission[26]. In addition, these quantum emitters can be deterministically coupled to plasmonic nanocavity gap modes by embedding WSe$_2$ between the nanocubes and a planar Au mirror, as depicted in **Figure 1a**. To avoid both optical quenching and spectral diffusion of the exciton emission the monolayer WSe$_2$ was separated from the Au surfaces by a 2 nm Al$_2$O$_3$ spacer layer. **Figure 1b** shows a scanning electron microscope image of monolayer WSe$_2$ strained over a 90 nm tall Au nanocube without piercing through. Photoluminescence (PL) mapping filtered over the exciton emission region (740–810 nm) demonstrates a high success rate of strain-induced exciton emission into the desired locations, as shown in **Figure 1c**. We have previously determined from time-resolved measurements that cavity-coupled quantum emitters created in this way reach deeply in the Purcell regime with average Purcell factors ($F_p$) over the array of $F_p$ =181. The Purcell effect results in an order of magnitude enhancement of the exciton QY from initially 1.5% in the uncoupled case to up to 12% for the coupled quantum emitters.

We first studied the uncoupled quantum emitters and found that the optical emission survives up to about 50 K, as shown in **Figure 2a**. These values are already larger than found in previous work (30 K[19,27]) for randomly occurring quantum emitters and are attributed to the energetically deeper exciton confinement potential that is induced by the gold nanopillars. When forming the coupled state by attaching the planar gold mirror we observe that the exciton emission survives up to about 80 K, as depicted in **Figure 2b.** This is a direct result of the Purcell-enhanced QY leading to a reduction in the thermally activated detrapping 0D-confined excitons. Note that these experiments in **Figures 2a-b** were carried out for monolayer WSe$_2$ exfoliated from a bulk crystal grown by the CVT technique. It was recently demonstrated that CVT growth of WSe$_2$ leads to a rather high density of point defects up to $10^{12}$ cm$^{-2}$ [30]. The underlying point defects such as metal vacancies and chalcogenide antiside defects can give rise to dominant $\tau_{NR}$, which is evident from the rather low QY of typically 1-2% in this material in the uncoupled state. To reach even higher operational temperatures for quantum emitters in WSe$_2$ we carried out comparable experiments with flux-grown crystals of WSe$_2$ that are characterized by a strongly reduced point defect density of typically 7×10$^{10}$ cm$^{-2}$ [30]. The corresponding spectra in **Figure 2c** show that the exciton emission can survive up to 160 K before merging into the noise floor, for the case that flux-grown WSe$_2$ is integrated into the plasmonic nanocavities. When scanning over the array at elevated temperatures, several quantum emitters can be found that emit at or above 160 K, with the best case reaching a record high value of 200 K.

Although exciton emission spectra can be recorded at elevated temperatures it is a priori not clear if the quantum light signatures can also survive to these elevated temperatures. Therefore, we record the second order photon correlation function g$^{(2)}$(τ) at 3.8 K base temperature as well as the highest accessible elevated temperature. As shown in **Figure 3a**, the correlation traces for the CVT-grown material show pronounced photon antibunching at zero-delay time with g$^{(2)}$(0) = 0.20 ± 0.03 at 3.8 K that degrades only slightly at 80 K to g$^{(2)}$(0) = 0.25 ± 0.04, which is still well below 0.5. Therefore, quantum light signatures are preserved even at temperatures allowing for device cooling with liquid nitrogen. Similarly, **Figure 3b** shows that for the case of flux-grown material g$^{(2)}$(0) = 0.22 ± 0.03 at 3.8 K while only slightly degrading at 160 K to g$^{(2)}$(0) = 0.35 ± 0.05. Depending on local variations over the quantum emitter array, we observed the highest temperatures for quantum light signatures g$^{(2)}$(0) < 0.5 ranging from 140 K to 180 K in flux-grown material (**Figure 3c**).

To better understand the contributions of the mechanisms responsible for thermal quenching of quantum emitters in monolayer WSe$_2$, we quantitatively analyze the exciton-phonon interaction as well as contributions from $E_A$ and $\tau_{NR}/\tau_R$ to the datasets in **Figure 2**. The temperature dependence of homogeneous linewidth for the emitters is shown in **Figure 4a**. For both material cases the excitation power was kept at a fixed value and thus the inhomogeneous linewidth broadening $\gamma_0$ due to pump induced dephasing and other effects should be unchanged. Under this condition the homogeneous linewidth broadening γ is dominated by thermally activated phonon scatterings according to:

$$\gamma = \gamma_0 + \sigma T + \gamma_{\text{LO}}(e^{\frac{\Theta}{kT}} - 1)^{-1}, \tag{1}$$

where $\sigma$ in the linear term is the acoustic phonon scattering coefficient. The exponential term describes thermal activation of optical phonons with energy Θ based on the temperature dependence of the Bose-Einstein distribution function. The corresponding fits shown by the solid lines in **Figure 4a** for CVT and flux-grown WSe$_2$ show only minor variation on the homogeneous broadening indicating that the exciton-phonon interaction is very similar for both cases. This is also expected since the change in point defect density predominantly affects $\tau_{NR}$, while the deformation potential scattering as well as the phonon density of states should remain unaffected.

In contrast to phonon scattering effects in the exciton linewidth, there is a clear difference between the two material cases regarding the spectral shift of the exciton emission with temperature, as illustrated in **Figure 4b**. The shift follows the band gap variation according to a Varshni-type equation[31]:

$$E(T) = E(0) - \alpha T^2/(T + \beta), \tag{2}$$

where $\alpha$ is related to the dilatation of the lattice and $\beta$ to the Debye temperature. For monolayer WSe$_2$ experimental values of $\beta$ = 170 K have been determined from Varshni analysis of exciton reflectivity spectra[32]. Using $\beta$ = 170 K for both cases in **Figure 4b**, the observed difference in energy shift fits to $\alpha_{\text{CVT}}$ = 1.8×10$^{-5}$ eV·K$^{-1}$ and $\alpha_{\text{Flux}}$ = 7.2×10$^{-6}$ eV·K$^{-1}$. It is apparent that the flux grown material, which is characterized by a factor 2.5 smaller α, shifts significantly less over the same temperature range as compared to the CVT grown sample. This can be understood as the underlying signature of larger local strain in the flux grown sample that traps the 0D exciton into an energetically deeper state. The presence of increased strain for the specific emitter studied in the flux-grown sample is also evident from the exciton transition energy that resides with 1.565 eV about 120 meV lower in energy as compared to the CVT grown sample. The deeper localization effectively decouples the exciton from the temperature dependent energy shift of the 2D bandgap.

The change of the local strain potential also directly leads to a change in $E_A$ that increases with increasing potential depth. To determine $E_A$ the Arrhenius plots in **Figure 4c.** displays PL intensity on a log-scale against inverse temperature. With increasing temperature, the intensity of the strain induced 0D excitons quench strongly with a characteristic extracted $E_A$ from the fit. Values between bare (49 ± 4 meV) and coupled quantum emitter (59 ± 5 meV) in the CVT grown sample vary within 20%. The activation energy of the 0D exciton in the flux grown sample in the coupled state is highest at values of 92 ± 4 meV, which is about 50% larger as compared to the CVT sample in the coupled state. These findings are in line with the observation of a smaller α parameter (**Figure 4b**) and a lower exciton transition energy in this case (**Figure 2**). We note that the given aspect ratio of the gold nanocube creates about 2% strain at the cube corners according to strain-profile simulations[26]. Since the aspect ratio was kept constant in this case the changes in the experimental $E_A$ values due to variation in the manually applied stamping pressure are rather moderate. To deliberately increase the strain depth, we furthermore studied pillars that are with 200 nm twice taller. In this case strain profiles up to 4% are predicted. However, the corresponding experimental activation energies are with values of $E_A$ =90 ± 2 meV (data not shown) quite comparable to the 100 nm tall pillars and indicate that the experiments do not reproduce this prediction. Most likely, the monolayer material slips over the substrate in the stamping process, thereby preventing creation of deeper exciton strain potentials in this approach.

In contrast, it is found that the survival of the 0D quantum emitters at elevated temperatures is predominantly affected by the material QY that affects the ratio $\tau_{NR}/\tau_R$ in the fit function. This ratio determines the turning point where the intensity drop accelerates, while $E_A$ modifies the slope. Using the QY values measured for the 0D excitons in our previous work for both CVT and flux grown material[26], we calculated the corresponding $\tau_{NR}/\tau_R$ values shown in **Table 1**.

As a direct consequence of the low QY of only 1.5% for bare quantum emitter in the CVT grown sample the PL intensity decreases sharply around 30 K and already drops at a temperature of 48.7 K below the 1/e value observed at 4 K. Coupling the quantum emitter in the CVT material to the plasmonic nanocavity mode drastically enhances the spontaneous emission rate by the Purcell effect, which leads to an 8-fold enhanced QY and $\tau_{NR}/\tau_R$ value that pushes the 1/e intensity point up to 66.6 K. The additional reduction of NR defect centers in the flux grown sample leads to an additional 6-fold reduction in $\tau_{NR}/\tau_R$ compared to the coupled CVT case, or 50-fold reduction compared to the bare case, that pushes the 1/e point up to 133.9 K, and allows recoding of optical spectra and antibunching traces up to 180 K. Apparently, the activation energy increase of about 50% contributes only as a minor effect to the quantum emitter survival, while the large drop in the $\tau_{NR}/\tau_R$ ratio is key for achieving high operation temperatures of the quantum emitters in WSe$_2$.

**Table 1. Fit parameters of the temperature dependent intensity of 0D excitons in WSe$_2$**

|  | Thermal activation energy (meV) | Quantum yield (%) | $\tau_{NR}/\tau_R$ | Temperature (K) at 1/e intensity value |
|---|---|---|---|---|
| Uncoupled CVT | 49 ± 4 | 1.5 | 65.7 | 48.7 |
| Coupled CVT | 59 ± 5 | 12 | 7.3 | 66.6 |
| Coupled Flux | 92 ± 4 | 44 | 1.3 | 133.9 |

Finally, we have calculated the minimum $E_A$ required to reach room temperature operation, as defined by an acceptable 1/e drop in exciton intensity at 300 K. For bare emitters in the CVT grown sample the required $E_A$ = 302 meV is quite high and would require at least an additional 6-fold improvement of the strain potential depth of the 0D exciton assuming a linear trend. Given that the Au nanocube already applies about 2 % strain to the monolayer, and the theoretical breaking point is around 6-11% with almost defect-free material[33,34], it may cause problems such as piercing of the monolayer material. In contrast, due to the largely improved QY, the minimum required $E_A$ for room temperature operation for the flux grown sample under coupling is reduced to 206 meV. As a result of reduced defect density combined with plasmonic nanocavity coupling, one requires only about a doubling of the demonstrated $E_A$=92 meV in this case.

In conclusion, we have demonstrated single photon emission from 0D quantum emitters in WSe$_2$ reaching up to 160 K. At these temperatures non-cryogenic cooling with thermo-electric devices becomes already feasible. To achieve room temperature operation in future work one would have to optimize the material growth to further decrease the NR defect density to increase QY from currently 64% in the best case[26] to about 98% under coupling. Particularly, when combined with active strain engineering through local actuators or through surface chemistry to reduce monolayer slippage, room temperature operation appears to be within reach. This is particularly interesting since the WSe$_2$ material platform appears to be currently the only system that allows spatial control of a large number of quantum emitters on a chip combined with deterministic coupling to plasmonic nanocavities[26].

## Methods

**Plasmonic chip fabrication:** The Au nanocube arrays were fabricated by an Elionix ELS-G100 electron-beam lithography (EBL) system using 495 Poly(methyl methacrylate) (PMMA) A4 (MicroChem) and developed in MIBK:IPA at a ratio of 1:3 for 180 sec. To form the 110 nm side-size nanocubes, a 5 nm Ti adhesion layer and a 90 nm Au layer were deposited on the samples with an electron beam evaporator (AJA Orion 3-TH). The remaining resist was then stripped with warm acetone at 50°C for 10 mins. We used atomic layer deposition (ALD) to add a 2 nm $Al_2O_3$ layer on top of the samples. To form the planar Au mirror, a 5 nm Ti adhesion layer and a 100 nm Au layer were deposited by slow-rate electron beam evaporation onto an epi-ready sapphire substrate.

**Material preparation:** The $WSe_2$ crystals for this study were synthesized based on the flux-growth technique by reacting W powder (99.999%) with Se shot (99.999%) in a ratio of 1:20. After loading these materials into a quartz ampoule together with a quartz wool scaffold the ampoule was sealed under a pressure of 1 mTorr and heated to 1000 °C over a period of 48 hours. The temperature was kept steady for 3 days before slowly cooling down to 400 °C, followed by centrifugation. Finally, the grown crystals were removed from the scaffold and annealed at 250 °C. Monolayers of $WSe_2$ were mechanically exfoliated either from commercial crystals (HQ Graphene Company) that were grown by the chemical vapor transport (CVT) technique or from the above described flux-grown crystals. The plasmonic chips were cleaned in Piranha solutions for 5 min and rinsed for 3 min in deionized water. For layer transfer we followed our previous dry stamping procedure utilizing an elevated substrate temperature of 60° C to prevent nanobubble formation[23].

**Photoluminescence spectroscopy:** Photoluminescence (PL) measurements were carried out using a closed-cycle cryogen-free cryostat (attodry1100, attocube systems AG) with variable temperature (3.8 K - 300 K). For optical excitation we utilized a laser diode operating at 532 nm in continuous wave mode. The $g^{(2)}(\tau)$ correlation traces were recorded with a Hanbury-Brown and Twiss setup and analyzed with a 4-channel time-to-digital converter (HRM-TDC, SensL).

## Acknowledgements

We like to thank Milan Begliarbekov for supporting the EBL process development at the City University of New York Advanced Science Research Center (ASRC) nanofabrication facility. S.S. acknowledges financial support by the National Science Foundation (NSF) under award DMR-1809235 and J.H. under award DMR-1809361. S.S. acknowledges financial support for the attodry1100 under NSF award ECCS-MRI-1531237.

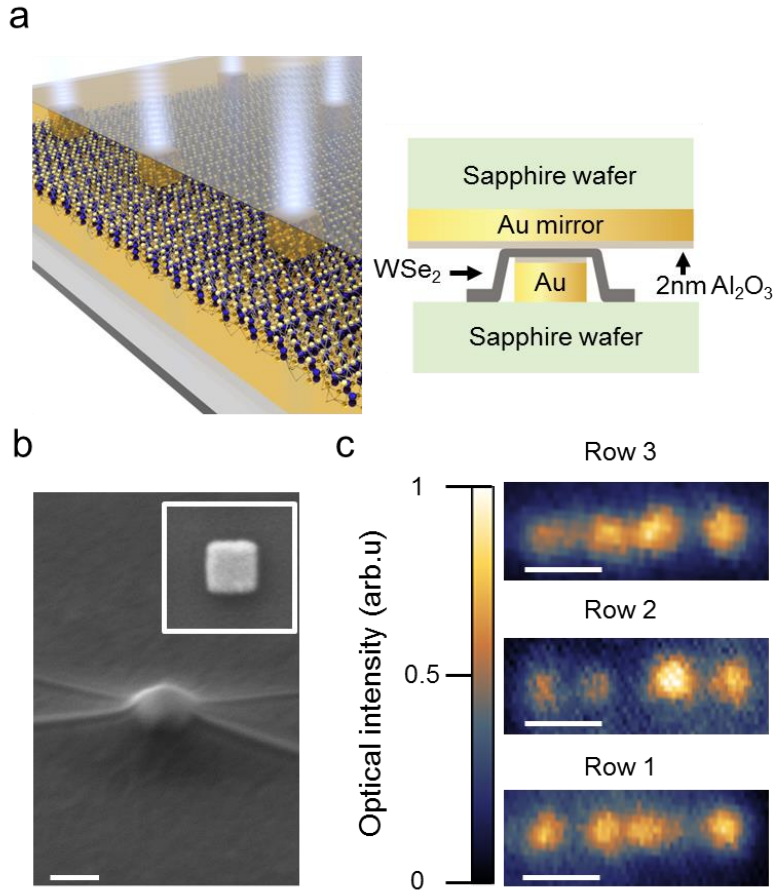

**Fig.1 Overview of sample design enabling deterministic coupling of strain-induced excitons to nanoplasmonic gap modes. a**, Left: Schematic of monolayer WSe$_2$ coupled to Au nanocube array. Right: Side view to show the layered design. The WSe$_2$ layer that is strained over the nanocube is separated by a 2 nm Al$_2$O$_3$ spacer on each side to avoid electric shortening with Au as well as optical quenching. The planar Au mirror can be separately attached to form the plasmonic gap mode. **b**, Scanning electron microscope (SEM) image of monolayer WSe$_2$ strained over Au nanocube viewing at 60 degree angle. Inset: SEM of Au nanocube before monolayer transfer. Scale bar: 100 nm. **c**, Photoluminescence maps of WSe$_2$ strained over Au nanocube array filtered within the spectral range 750-850 nm. Scale bar: 2 μm.

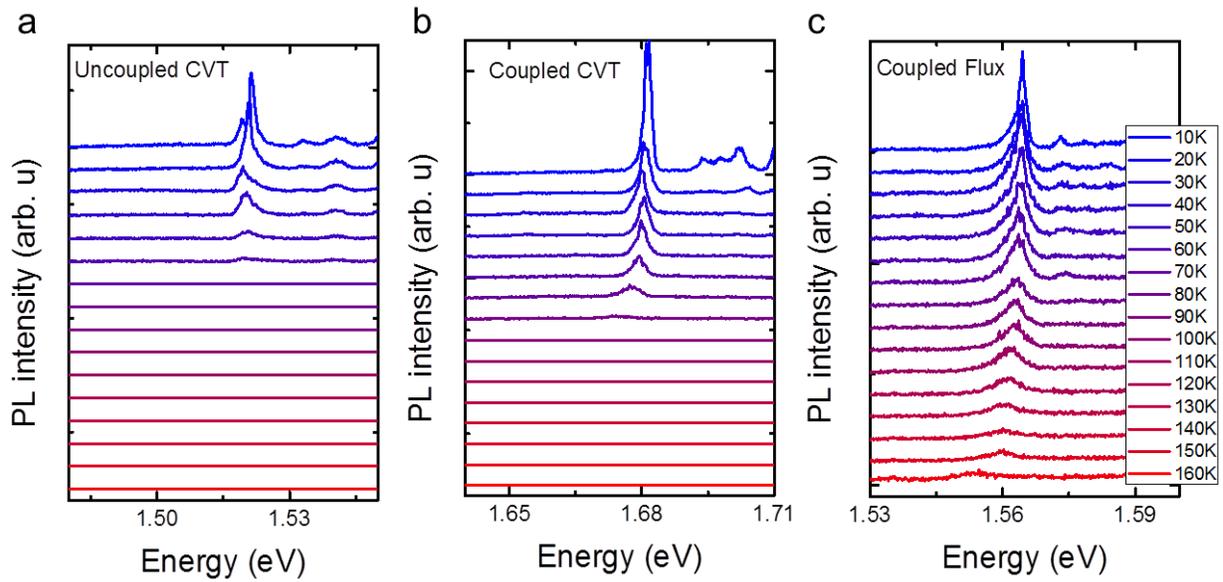

**Fig.2 Temperature-dependent photoluminescence (PL) spectra of strain-induced quantum emitters.**
**a,** PL spectra recorded in the temperature range from 10 K to 160 K for a monolayer of CVT-grown $WSe_2$ that is stamped onto the gold nanocube array without attaching the planar mirror **b,** Same for the case that the planar mirror is attached defining the coupled state that leads to a Purcell-enhanced quantum yield **c,** Strain-induced quantum emitter in the coupled state based on a monolayer of flux-grown $WSe_2$. Data are offset vertically for clarity.

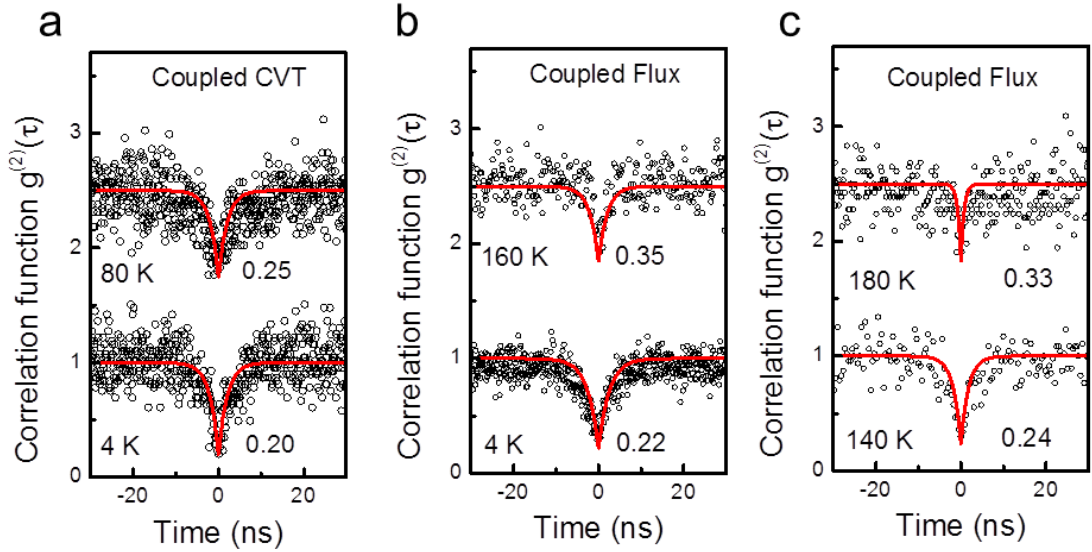

**Fig.3 Quantum light emission from cavity-coupled quantum emitters at elevated temperatures.** Open black dots are second-order photon correlation traces $g^{(2)}(\tau)$ recorded with a Hanbury-Brown Twiss interferometer under non-resonant excitation. The red solid lines are fits to the rate-equation analysis for an individual two-level system. **a,** Exemplary quantum emitter from CVT-grown material resulting in single-photon purity values of $g^{(2)}(0) = 0.20 \pm 0.03$ at 4 K and $g^{(2)}(0) = 0.25 \pm 0.04$ at 80 K. **b,** Corresponding traces for an exemplary quantum emitter from flux-grown WSe$_2$ resulting in a single-photon purity values of $g^{(2)}(0) = 0.22 \pm 0.04$ at 4 K and $g^{(2)}(0) = 0.35 \pm 0.05$ at 160 K. **c,** Two additional cases for quantum emitters in flux-grown WSe$_2$ showing single-photon purity values of $g^{(2)}(0) = 0.24 \pm 0.03$ at 140 K and $g^{(2)}(0) = 0.33 \pm 0.08$ at 180 K.

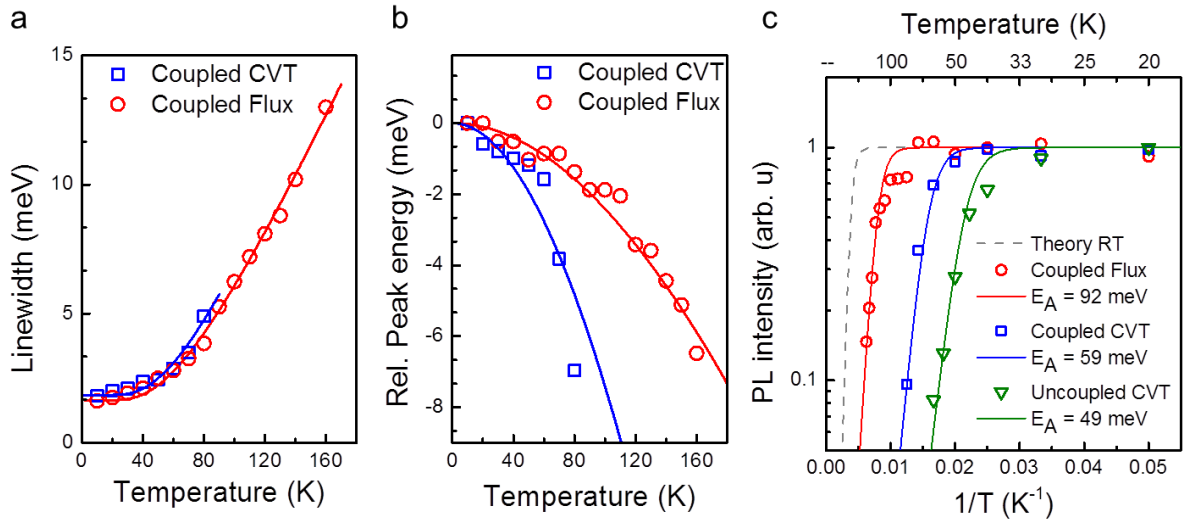

**Fig. 4 Temperature dependence of the spectral linewidth, emission energy, and intensity of the exciton emission. a,** Temperature dependence of the spectral linewidths comparing coupled CVT-grown (blue squares) and coupled flux-grown (red circles) quantum emitters. Solid lines are analytical fits considering acoustical and optical phonon scattering **b,** Exciton emission energy measured relative to the value at 4K (relative peak energy) comparing CVT-grown quantum emitter (blue squares) with flux-grown quantum emitter (red circles). Data are fitted using the Varshni relation (solid lines). **c**, Arrhenius plot of integrated PL intensities recorded in the temperature range from 20 K to 180 K for uncoupled (green triangles) and coupled (blue squares) with CVT grown $WSe_2$ and coupled case with flux grown $WSe_2$ (red circles). Solid lines correspond to theoretical fits using the various expressions given in the text. Grey dash line shows the theoretical calculation for coupled flux case to reach room-temperature operation. All PL spectra are taken under 60 µW excitation power.